\documentclass[useAMS,usenatbib]{mn2e}

\usepackage{amssymb, amsmath, amsfonts}

\usepackage{graphicx, shadow}

\usepackage{epstopdf}

\usepackage{rotating}
\usepackage{wasysym}
\usepackage{verbatim}
\usepackage{natbib}
\usepackage{hyperref}

\newcommand{\sunrise}{\textsc{Sunrise}}

\newcommand{\hst}{{\sl HST}}

\newcommand{\hubble}{{\sl Hubble Space Telescope}}
\newcommand{\webb}{{\sl James Webb Space Telescope}}

\newcommand{\euclid}{{\sl Euclid}}
\newcommand{\wfirstfull}{{\sl Wide Field Infrared Survey Telescope}}

\newcommand{\candels}{{CANDELS}}

\newcommand{\illustris}{{Illustris}}

\renewcommand{\vec}[1]{ {\bmath #1} }

\newcommand{\bigbox}{$\left (106.5\rm\ Mpc\right )^3$}

\title[]{Massive Close Pairs Measure Rapid Galaxy Assembly in Mergers at High Redshift}

\author[Modeling Close Pairs in Illustris]{Gregory F. Snyder$^1$\thanks{Giacconi Fellow}, Jennifer M. Lotz$^1$, Vicente Rodriguez-Gomez$^2$, \newauthor Renato da Silva Guimar\~{a}es$^{1,3,4}$, Paul Torrey$^4$\thanks{Hubble Fellow}, Lars Hernquist$^5$ \\ 
$^1$ Space Telescope Science Institute, 3700 San Martin Dr, Baltimore, MD 21218 \\
$^2$ Department of Physics \& Astronomy, Johns Hopkins University, 3400 N. Charles St, Baltimore, MD 21218, USA \\
$^3$ Instituto de F\'{i}sica, Universidade de S\~{a}o Paolo, Rua do Mat\~{a}o 1371, Butant\~{a}, S\~{a}o Paulo, 05508-090, Brazil \\
$^4$ Department of Physics, Kavli Institute for Astrophysics \& Space Research, Massachusetts Institute of Technology, Cambridge, MA, 02139, USA \\
$^5$ Harvard-Smithsonian Center for Astrophysics, 60 Garden St, Cambridge, MA, 02138, USA \\
}
\begin{document}


\maketitle

\begin{abstract}

We compare mass-selected close pairs at $z > 1$ with the intrinsic galaxy merger rate in the Illustris Simulations. To do so, we construct three 140 arcmin$^2$ lightcone catalogs and measure pair fractions, finding that they change little or decrease with increasing redshift at $z > 1$.  Consistent with current surveys, this trend requires a decrease in the merger-pair observability time, roughly as $\tau \propto (1 + z)^{-2}$, in order to measure the merger rates of the same galaxies.  This implies that major mergers are more common at high redshift than implied by the simplest arguments assuming a constant observability time.  Several effects contribute to this trend:  (1) The fraction of massive, major (4:1) pairs which merge by today increases weakly from $\sim 0.5$ at $z=1$ to $\sim 0.8$ at $z=3$.  (2) The median time elapsed between an observed pair and final remnant decreases by a factor of two from $z\sim 1$ to $z\sim 3$.   (3) An increasing specific star formation rate (sSFR) decreases the time during which common stellar-mass based pair selection criteria could identify the mergers.  The average orbit of the pairs at observation time varies only weakly, suggesting that the dynamical time is not varying enough to account by itself for the pair fraction trends.  Merging pairs reside in dense regions, having overdensity $\delta \sim 10$ to $\sim 100$ times greater than the average massive galaxy.  We forward model the pairs to reconstruct the merger remnant production rate, showing that it is consistent with a rapid increase in galaxy merger rates at $z > 1$.

\end{abstract}


\section{Introduction} \label{s:intro}

In order to disentangle the various mechanisms driving galaxy formation, we must measure accurately the rate of mergers between galaxies over cosmic time. Simulations of $\Lambda$CDM universes using both N-body and hydrodynamical methods predict a decreasing galaxy major merger rate with cosmic time \citep[e.g.,][hereafter R-G15]{Fakhouri2008,Rodriguez-Gomez2015}.  Surveys of galaxies at $z \lesssim 1$ have found that the incidence of massive, major pairs decreased rapidly as the universe evolved \citep[e.g.,][]{Kartaltepe2007,lin08}.  Assuming a roughly constant observability timescale, i.e., the duration a merging pair is identified in a catalog, these observations of galaxy pairs matched well with expectations \citep[e.g.,][]{hop10b,Lotz2011,Robotham2014}. {Moreover, different windows on the merger process, such as asymmetric morphologies \citep{Conselice2003} and post-starburst signatures \citep{snyder11a} all generally agree with this picture, so long as we adopt the proper observability timescales.}

However, recent surveys found that the incidence of massive, mass-ratio-selected major pairs \emph{remains constant or turns over} at early times \citep{Ryan2008,Williams2011,Man2012,Man2016,Ferreras2013}, apparently contradicting the simplest expectations from theory.  Possible reasons include a true decline in the incidence of mergers in galaxies at early times, incompleteness, and changes in merger properties. While these results were originally somewhat uncertain, multiple groups have now reached agreement about this conclusion \citep[e.g.,][Mantha et al. in prep.]{Mantha2016AAS}. Other selections, such as late-stage morphological disturbance \citep[e.g.,][]{Bluck2012} and major flux-ratio pairs \citep[e.g.,][]{Man2016}, show the expectedly increasing merger fraction at $z \gtrsim 2$, but these selections likely include a large number of minor mergers with stellar mass ratio less than 1 to 4 \citep[e.g.,][]{Lotz2011}.   

{  
Some previous studies of galaxy merger simulations \citep[e.g.,][]{lotz08,lotz10,Lotz2010} found that the ensemble-averaged observability time of major merger pairs is roughly constant at $\tau_{\rm obs} \sim 0.5$ Gyr. An assumption like this one underlies many of the observational estimates of the galaxy merger rate at high redshift. However, this timescale value was derived from suites of idealized merger simulations and coarsely resolved N-body plus semi-analytic models tailored to observations at $z \lesssim 1.5$. Therefore, it is not necessarily correct to apply this timescale to new data at high redshift in order to infer evolution in the galaxy merger rate. For example, merger events could proceed differently at early times and in different large-scale environments owing to variations in local density or galaxy contents. These changes, if present, would represent evolution in the merger dynamical timescales. By itself, a varying dynamical timescale will not affect the \emph{observability time} required to translate observed pairs into a merger rate. However, because we do not know how mergers or their observability times evolve in all situations, we must be cautious when assuming these values. 

Moreover, when comparing observed estimates of merger rates to theoretical expectations, it is critical that the observability timescales we apply account for all possible processes affecting the detectability of mergers.  While high-resolution hydrodynamical simulations allow us to model the evolution of galaxies relatively accurately during mergers, their expense often prevents us from simultaneously modeling their full cosmological history, including environmental effects and the full diversity of mass accretion histories \citep[c.f.,][]{Pedrosa2014,Snyder2015a,Bignone2017}. In principle, this cosmological context could drive evolution in observable signatures important for selecting pairs, such as masses, luminosities, velocities, and clustering. Therefore, the average observability times we must assume to measure intrinsic merger rates could depend on redshift, density, or other variables. These complications motivate us to continue augmenting estimates of merger observability timescales by basing them on systematic, detailed simulations in a cosmological context.

The recent emergence of relatively large ($\sim 100\rm\ Mpc$) hydrodynamical simulations with moderate resolution ($\sim 1\rm\ kpc$) and accuracy offers an opportunity to refine our assumptions about galaxy merger observability timescales \citep[e.g.,][]{Dubois2014,Khandai2015,Schaye2014,Vogelsberger2014b}.  Their cosmological sizes allow us to convert these simulations into mock catalogs in the same way as has been done for several years using semi-analytic and semi-empirical models, for example by using the lightcone technique to create mock galaxy catalogs as a function of angular coordinates and redshift \citep[e.g.,][]{Blaizot2005,Kitzbichler2007,Carlson2010,Bernyk2016,Overzier2012,Taghizadeh-Popp2015}.  This has the benefit of bringing our comparisons between theory and data into the same language. Moreover, by further applying stellar population, dust modeling, and instrument simulation, we can fully forward model these simulations into fluxes per pixel, capturing not only large-scale structure but also internal galaxy dynamics as we would measure them. These mock data can then be analyzed using processing pipelines identical to surveys. Coincidentally, the areas and spatial resolution probed by deep galaxy surveys such as CANDELS \citep{Koekemoer2011,Grogin2011} and 3D-HST \citep{Brammer2012,Skelton2014} are roughly similar to what we can achieve using lightcones created from the best available hydrodynamical models.  Therefore, these simulations are well-suited to deriving improvements in the mapping between observed and intrinsic galaxy mergers via new estimates of observability timescales.

A possible downside to this approach is that our theoretical catalogs will have the same limitations inherent to our observational perspective, including limited statistics at low redshift in synthetic pencil-beam surveys. The lightcone technique may impart spurious correlations to widely separated objects in the mock survey, owing to the spatial sampling algorithm or limitations in simulation volume. Simulation-based mock catalogs will also depend sensitively on the particular set of physical assumptions underlying the model. For example, the galaxies simulated by any given model are not yet perfect matches to observed summary statistics. Moreover, the tools used to identify substructure, which we use to define and label galaxy mergers, are also imperfect in certain relevant situations where the assignment of mass might be ambiguous (see discussion by R-G15). Therefore, it is important to keep these limitations in mind when making comparisons and using timescales derived in this fashion.

Another advantage to using lightcone models is that we can define mergers in theory and data space using nearly identical selections. Often, we measure the galaxy merger rate in simulations as a function of mass and time using an idealized method for counting merger events, a method that observations of pairs cannot access.  Also, we can directly disentangle certain observational effects such as how often a particular pair criterion selects a false positive--- galaxies which are physically unrelated.  Finally, when making comparisons, a common flaw is to plot the evolution of observational merger fractions (sometimes measured in incompatible ways) versus redshift, but compare only to simulated merger rates measured in the idealized fashion. By converting large, detailed simulations into an observational frame, we can dispense with this limitation and contrast directly our observational selections with our theoretical understanding.
}

In this paper, we model surveys of massive, mass-ratio-selected major galaxy pairs in the \illustris\ Simulations, and recover evolution in the mock-observed pair fractions consistent with the perplexing observations of a flat merger fraction at $z > 1$. Because the intrinsic galaxy merger rate in \illustris\ declines rapidly versus cosmic time, the average time over which these pairs can be observed must increase commensurately. 

We describe the construction of our synthetic survey catalogs in Section~\ref{s:methods}.  We present the basic pair statistics and merger rate synthesis in Section~\ref{s:pairs}, comparing to observational data in Section~\ref{ss:datacompare}.  We demonstrate how the properties of simulated pairs evolve in such a way as to yield the observed trends in Section~\ref{s:properties}.  We summarize and discuss future directions with mock data in Section~\ref{s:summary}.

\section{Methods} \label{s:methods}

To best inform the identification of mergers in survey data, our goal is to create synthetic survey fields from models of galaxy formation, which has been achieved with semi-analytic \citep[e.g.,][hereafter O13]{Bernyk2016,Overzier2012} and semi-empirical models \citep[e.g.,][]{Taghizadeh-Popp2015}, as well as hybrids of these with the results of hydrodynamical simulations \citep{Hayward2013}. In principle, any galaxy formation model sufficiently large, and detailed enough to track subhalos in the mass range of interest, would support these calculations. While the present focus is on investigating mergers via pair statistics, our long-term goal includes using detailed morphological information which is available on $\sim$kpc scales at high redshift in the best imaging.  Because full snapshot data were public at the initiation of this project \citep{Nelson2015DR}\footnote{available: \href{http://www.illustris-project.org/data}{www.illustris-project.org/data} }, we chose the \illustris\ Simulations \citep{Vogelsberger2014b,Vogelsberger2014a,Genel2014} for our galaxy survey field models.  

The \illustris\ Project consists of hydrodynamical simulations of galaxy formation in $(106.5\rm Mpc)^3$. Using the Arepo code \citep{Springel2010}, \illustris\ applied galaxy physics consisting of cooling, star formation (SF), gas recycling, metal enrichment, supermassive black hole (SMBH) growth, and gas heating by feedback from supernovae and SMBHs. Precise models were chosen to match the $z=0$ stellar mass and halo occupation functions, plus the cosmic history of SFR density \citep{Vogelsberger2013}. This reproduces the observed stellar mass function, SFR-Mass main sequence, and Tully-Fisher relation, from z=3 to z=0 \citep{Torrey2014}. \citet{Genel2014} highlighted successes with satellite number densities, baryon content versus radius, and specific SFRs up to $z = 8$.  \citet{Rodriguez-Gomez2015} showed that the galaxy merger rate compares favorably to observations at $z < 1$ and predicted its evolution as a function of time, mass, and mass ratio.  In this paper, we present results from the Illustris-1 simulation (the highest resolution) but have confirmed that our results are qualitatively unchanged when using Illustris-2 catalogs instead.

To generate mock surveys, we take into account that the simulation consists of a cubic fixed comoving volume $L^3$ with periodic boundary conditions.  In brief, creating a mock survey involves remapping the cubic simulation volume to a shape that preserves large-scale structure information but can more accurately mimic observation geometries.  Several techniques exist \citep{Blaizot2005, Kitzbichler2007, Carlson2010}, either with a single simulation output or replacing distant volume with the output from an earlier cosmic time.  This latter approach defines lightcones in which the properties of galaxies evolve roughly as they would be observed in surveys.  

\subsection{Lightcone geometries} \label{ss:geometry}

To create lightcones, we follow \citet{Kitzbichler2007}, in which we choose two integers $(n,m)$ that set the viewing direction from the origin of the cubic volume.  The viewing direction is defined by the unit vector $\hat{u}_3$: \[ \hat{u}_3 = (n,m,nm)/| (n, m, nm) |. \]  For simplicity, we restrict our fields to a square geometry and set the angular width of the theoretical cameras to the smaller of the possible dimensions: \[ \frac{1}{m n^2}\rm\ radians. \]    

We trace a line along $\hat{u}_3$ until it exits the box (the $z=0$ output) and enters the next periodic replication, and repeat this until we reach the final desired comoving distance.  If the comoving distance traveled is such that output exists at a redshift appropriate for the distance to the midpoint of the next segment, we use this new output (from an earlier cosmic time) to fill the new segment.  The viewing direction vector and angular field of view are such that the lightcone will trace out a unique comoving volume up to $nm$ replications.  This unique volume is $V_0 \approx L^3 / 3$.   This and similar procedures have been applied in many previous modeling analyses \citep[e.g.,][]{Blaizot2005,Kitzbichler2007,Henriques2012,Bluck2012,Overzier2012,Taghizadeh-Popp2015,Bernyk2016}.  

For \illustris, \bigbox, choosing $(n,m)=(11,10)$ produces a lightcone subtending a unique volume from $z=0$ to $z \sim 18$ over $\sim 8\rm arcmin^2$.  We refer to this setting as our Thin lightcones.  It is possible to make slightly larger mock surveys by permitting some repetition.  For example, setting $(n,m) = (7,6)$ yields a $\sim 140\rm\ arcmin^2$ survey field, roughly equivalent to the total ``ultra deep'' cosmological imaging by \hst.  We refer to this setting as our Wide lightcones.  

The Wide setting yields a $11.7 \times 11.7$ arcmin$^2$ field containing unique comoving volume from $z=0$ through $z \approx 1.57$.  Above this redshift, the lightcone segments repeat comoving space that was passed in the early segments (near $z\sim 0$) but at very different epochs ($z \sim 2$ and $z \sim 9$).  Because the beam was so narrow and the time so different, this repetition is difficult to notice.  More obvious is that above $z = 1.57$, adjacent segments will begin to repeat the same comoving space at very similar cosmic times near the edges of the lightcone field.  This results in a loss of statistical power from what would be expected in a true astronomical survey of these dimensions, starting with a very small effect at $z \sim 2$ but roughly halving the truly unique area by $z \sim 4$.  This yields a benefit in that we are more likely to be sampling a significant fraction of the simulation volume at $z \sim 2$ in each lightcone, but this is a tradeoff against somewhat poorer and more correlated statistics than a larger simulation would enable.  

For each of the Thin and Wide geometry, we denote our default lightcone ($\vec{u}_3 = (n, m, nm)$) as Field B and create two additional fields by swapping coordinates: x with y to create Field A, and x with z to create Field C.  These three fields do not necessarily contain unique objects, but the 3D structure will be viewed from different directions.

\subsection{Catalog synthesis and pair counting}  \label{ss:catalogs}

Our Wide geometry subtends a square $11.7$ arcmin per side, $\sim 136$ arcmin$^2$, larger than the CANDELS-Deep survey area. The three Wide fields together total $\sim 410$ arcmin$^2$, about half the area surveyed by CANDELS-Wide \citep{Grogin2011,Koekemoer2011}. Each Thin lightcone is $\sim 8$ arcmin$^2$, a little smaller than the \hubble\ Ultra Deep Field \citep[$11$ arcmin$^2$][]{Beckwith2006}, and we used these to create the synthetic Deep Fields for \illustris\ \citep{Vogelsberger2014a}. For our investigation into pair statistics in this Section, we use the Wide lightcones.

With these pre-defined lightcone geometries, we use the \illustris\ Data Release \citep{Nelson2015DR} example scripts to parse the galaxy catalogs and select those with baryonic mass above $10^{9.5} M_{\odot}$ residing in the lightcone volume.  We assign them angular coordinates, comoving distances, and total redshifts by projecting their positions and velocities from simulation space into the frame of a hypothetical observer at the origin.  We transfer all galaxy properties into a catalog and then choose pairs using selection techniques that are commonly applied to real survey catalogs.

We have explored numerous selection criteria, but we will focus on the following selection throughout much of this paper:
\begin{enumerate}
\item{The primary source has a stellar mass $10.5 < \log_{10} M_{\rm pri}/M_{\odot} \le 11.0$. }
\item{The secondary source has a stellar mass $0.25 M_{\rm pri} \le M_{\rm sec} \le M_{\rm pri}$.  In other words we use a 4:1 mass ratio selection. }
\item{The two sources have a projected distance $10\rm kpc/h$$ \le d_{\rm proj} \le  $ $50\rm kpc/h$ measured at the location of the primary with redshift $z_{\rm pri}$.  }
\item{The two sources have a relative redshift $\Delta z = |\Delta v|/c  \le 0.02 (1+z_{\rm pri})$, corresponding to a maximum offset velocity of $18000\rm\ km/s$ at $z=2$, roughly matching the photometric redshift uncertainty for massive galaxies in deep surveys. }
\end{enumerate}

Because rates can be a strong function of mass, we use an upper mass limit in item i to reduce confusion about the contribution from very massive galaxies.  { When removing this upper mass limit, our conclusions are unchanged, because galaxies near the lower selection cut dominate all pair and primary samples owing to the steep mass function in massive galaxies.  }

Item ii deserves careful consideration: the ratio between the stellar mass of the secondary and primary in the pair (hereafter, ``mass ratio'').  In our lightcone catalogs, these masses derive from the {\sc Subfind} code \citep{Springel2001,Dolag2009} used to assign cosmic structure to a galaxy catalog.  The assigned masses are known to depend on the specific algorithm used to measure substructure, and {Subfind} suffers from a tendency to improperly assign masses between halos experiencing a merger.  This can be problematic for measuring the intrinsic merger rates, and motivated R-G15 to define a new mass ratio at the timestep ($t_{\rm max}$) when the secondary galaxy achieved its maximum stellar mass.  For the case of true mergers in a single simulation volume, this $t_{\rm max}$ solution enabled a robust measurement of the intrinsic merger rates as a function of mass, time, and mass ratio.  However, the $t_{\rm max}$ method is not necessarily consistent with survey mass ratio estimators, because we cannot measure the prior or future mass history in observed samples.  Data are therefore limited to measuring mass ratios at the time observed ($t_{\rm obs}$), and these masses can also be affected by mass stripping or evolution triggered by merger or environmental processes themselves.

The ideal solution to this challenge is to project all intrinsic baryonic mass elements separately and create full synthetic mock datasets, such as survey imaging mosaics, for example as has been done by the Millennium Run Observatory (O13).  For simplicity, in this paper we restrict ourselves to pair measurements based on simulated catalogs, and confirm that our main conclusions are the same whether we use a $t_{\rm obs}$ or $t_{\rm max}$ selection in Section~\ref{ss:massratiodef}.  

In this paper, we denote points in cosmic time as $t$, and durations as $\tau$.  For example, we define: 
\begin{align*}
\tau_{\rm obs} &: \rm observability\ time; \text{the period during which} \\ 
&\ \ \text{the pair could be selected.} \\
\tau_{\rm merge} &: \rm merger\ duration\ or\ \text{dynamical time} \\
t_{\rm obs} &: \rm cosmic\ time\ we\ observed\ the\ pair \\
t_{\rm merger} &: \rm time\ that\ merger\ occurs. 
\end{align*}

Other definitions we will use include:  
\begin{align*}
R_m &: \rm number\ of\ mergers\ per\ galaxy\ per\ Gyr \\
\tau^{\rm SFR}_{\rm obs} &: \rm duration\ pair\ satisfies\ mass\ selection\ cuts.
\end{align*}

\section{Pair Statistics and Merger Rates}   \label{s:pairs}
In this section, we measure the pair fractions as a function of redshift in the Wide lightcone catalogs and compare against the intrinsic galaxy merger rates from R-G15.  Each of the three Wide fields is $\sim 140$ arcmin$^2$, totaling $\sim 410$ arcmin$^2$, or over half of the total area of the \candels\ \hst\ survey \citep{Koekemoer2011,Grogin2011}. Thus our models are well suited to existing measurements from subsets of the \candels\ data.  

\begin{figure}
\begin{center}
\includegraphics[width=3.5in]{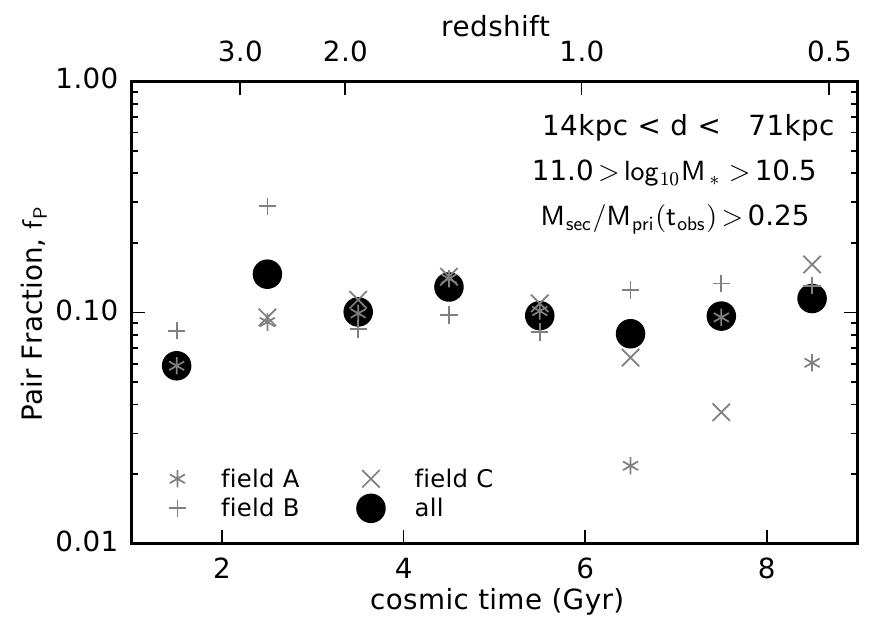}
\caption{Massive, major pair fractions versus cosmic time for our three Wide lightcone catalogs from Illustris.  We select pairs to have a relative line-of-sight redshift difference $|\Delta v|/c < 0.02*(1 + z)$, which is $|\Delta v| < 18000$ km/s at $z=2$, comparable to photometric redshift precision in deep surveys. {For the highest redshift point in Field C, the pair fraction is zero.} \label{fig:pairs}}
\end{center}
\end{figure}

Figure~\ref{fig:pairs} presents the pair fractions versus cosmic time.  At $z > 1$, our three fields each contain a very similar number of pairs, indicating that we are not cosmic-variance limited.  With our default selections, pairs constitute about $10\%$ by number of the massive galaxy sample, with very little or no evolution from $z=1$ to $z=3$.  This result is a fair match to observations \citep[e.g.,][]{Williams2011,Man2016} which find very little, no, or even negative evolution in pair counts from $z=1$ to $z=3$.  Moreover, a recent analysis of the Eagle Simulation found a similar flattening in theoretical pair fractions at high redshift \citep{Qu2016}, corroborating our results.  

\begin{figure}
\begin{center}
\includegraphics[width=3.5in]{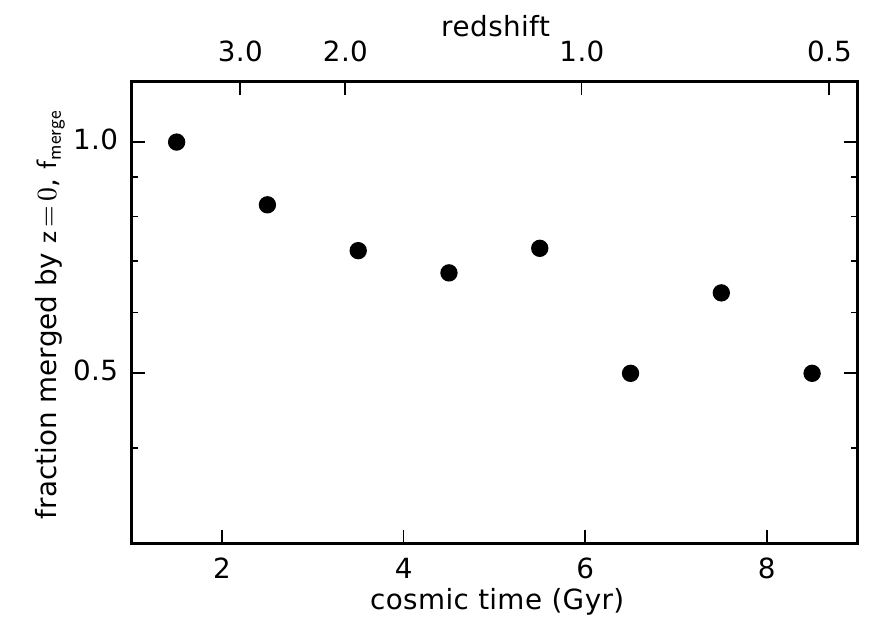}
\caption{Fraction of pairs selected in Figure~\ref{fig:pairs} which merge by $z=0$. At $z \sim 3$, we find that the fraction of false positives --- massive pairs which do not merge by $z=0 $ --- is approximately $20\%$, while at $z \sim 1$ it is $\sim 50\%$. \label{fig:mergers}}
\end{center}
\end{figure}

As we show in Figure~\ref{fig:mergers}, the pair samples we select are very pure at $z > 1$. The fraction of massive pairs selected in Figure~\ref{fig:pairs} that are true mergers decreases from $\sim 80\%$ at $z > 1$ to $\sim 50\%$ by $z = 0.5$.  At $z \lesssim 1$, we find severe cosmic variance owing to the small volume surveyed, and so we cannot make strong conclusions about how these lightcone-based pair fractions evolve to $z=0$.  With larger simulations, we would expect to recover the expected decrease in the merger fraction from $z=1$ to $z=0$ \citep[e.g.,][]{Lotz2011,Man2016}.

\subsection{Merger rates} \label{ss:mergerrates}

\begin{figure}
\begin{center}
\includegraphics[width=3.5in]{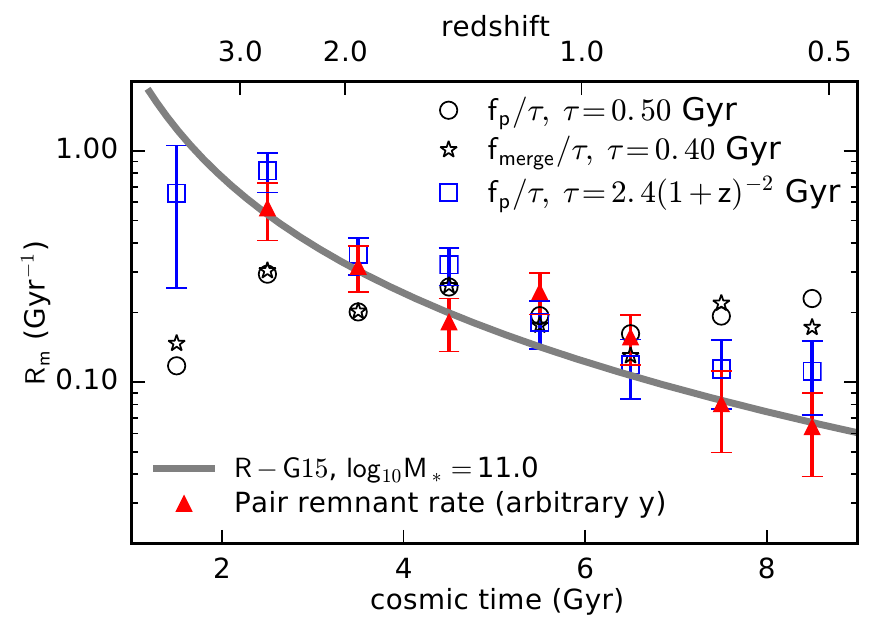}
\caption{Number of mergers per galaxy per Gyr versus cosmic time inferred from galaxy pair statistics.  Black circles assume a constant observability timescale $\tau = 0.5\rm\ Gyr$, blue squares assume that $\tau = 2.4 (1+z)^{-2}\rm\ Gyr$, and stars ignore false positives and assume $\tau = 0.4\rm\ Gyr$.  { Red triangles show a reconstruction of the remnant formation rate implied by the mock-observed pairs, described in Section~\ref{ss:remtimes}.} Only the blue squares and red triangles adequately match the evolution in merger rates measured by \citet{Rodriguez-Gomez2015}, and therefore we infer a rapidly evolving pair observability timescale $\tau$. \label{fig:rates} }
\end{center}
\end{figure}

To infer the galaxy merger rate from pairs, we must assume or measure the merger-pair observability time $\tau$ corresponding to the duration that a merger can be selected as a given pair.  However, we also know the intrinsic merger rates \emph{a priori} because they have been measured by R-G15.  We compare these approaches in Figure~\ref{fig:rates}.  

The default selection has a low false positive rate in that it selects pairs of which $\sim 80\%$ are truly close at $z \sim 2$ and ultimately merge (Figure~\ref{fig:mergers}).  This is also demonstrated by the small difference in trends between open circles (total pair counts divided by $0.5$ Gyr) and open stars (true merging pairs divided by $0.4$ Gyr) in Figure~\ref{fig:rates}.  This implies that the pair fraction trend we measure is driven by real merging pairs and it is not an effect caused by unrelated chance projections.  It also implies that all else being equal, photometric redshifts with our assumed precision of $0.02 (1+z)$ should be sufficient to capture a relatively pure sample of merging pairs at $z > 1$.  

The flat pair fractions trace a trend rather different from the intrinsic merger rates given by R-G15, which rise rapidly at early times.  Assuming a constant $\tau$, neither the pair nor merger fraction trends match adequately to the true merger rate evolution in the simulation: assuming our optimistic normalization procedure, the binned curves differ by $8.1\sigma$ and $7.4\sigma$, respectively (reduced $\chi^2_r$ of $8.2$ and $6.8$).  

We argue that an evolving merger pair observability time $\tau$ could explain this discrepancy.  In Figure~\ref{fig:rates}, we find that assuming $\tau = 2.4 (1+z)^{-2}$ Gyr results in an adequate match between the pair-based and intrinsic merger rates ($\chi^2_r \approx 1.1$).  

{ 
Moreover, in Section~\ref{ss:remtimes}, we will discuss an approach to forward-model the mock-observed pairs, using their known merger completion time to derive a remnant formation rate versus time.  We demonstrate this approach in Figure~\ref{fig:rates} with red triangles, which also adequately trace the intrinsic merger rates derived by R-G15.
}

\subsection{Effect of mass ratio definition} \label{ss:massratiodef}

\begin{figure}
\begin{center}
\includegraphics[width=3.5in]{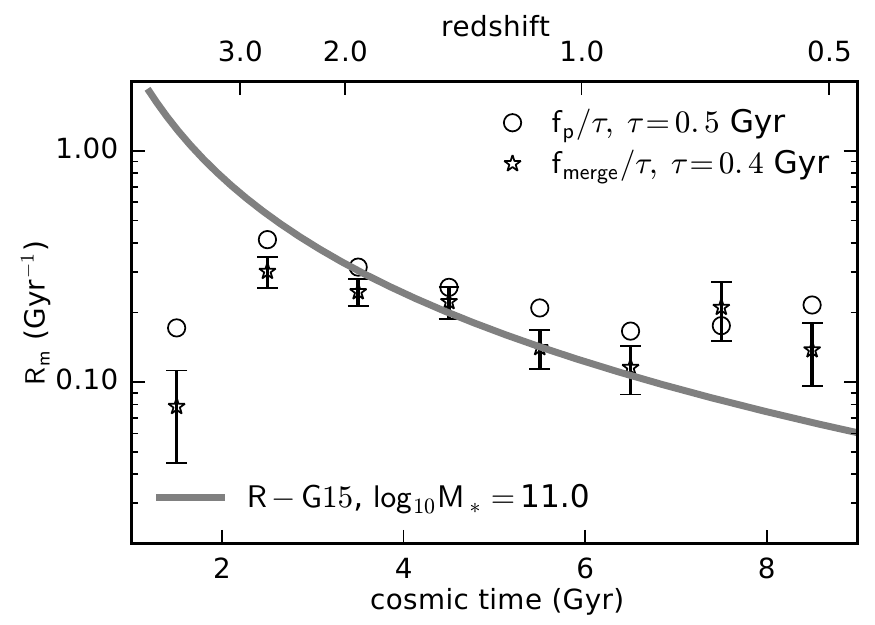}
\caption{Similar to Figure~\ref{fig:rates} except using the $t_{\rm max}$ mass ratio definition from \citet{Rodriguez-Gomez2015}, and showing fewer data series for clarity.  This method selects roughly the same true mergers at $z \sim 1$-$2$, slightly more true mergers at $z \sim 3$, and roughly a factor of two more false projections overall compared with the $t_{\rm obs}$ definition.  We infer a correspondingly weaker trend in observability times: although the pair statistics are still consistent with a $\tau \propto (1+z)^{-2}$ trend, the goodness of fit in the open circles and stars is more acceptable without this strong evolution.  \label{fig:rates_tmax} }
\end{center}
\end{figure}

So far, we have been using pair fractions measured with a mass ratio defined at $t_{\rm obs}$ to compare against merger rates measured with a mass ratio defined at $t_{\rm max}$ by R-G15.  Some, but not all, of the differences above result from our choice of mass ratio definitions, as we show in Figure~\ref{fig:rates_tmax}.  Switching from a $t_{\rm obs}$ to a $t_{\rm max}$ mass ratio definition (see Section~\ref{ss:catalogs}) has two main effects:
\begin{enumerate}
\item{The rate of false projections is significantly higher with the $t_{\rm max}$ definition, comprising roughly half of the pair sample.  }
\item{Slightly more true major mergers are identified at $z \gtrsim 2$ using $t_{\rm max}$ rather than $t_{\rm obs}$.  }
\end{enumerate}

The first effect results from the need to define a $t_{\rm max}$ mass ratio even for pairs that don't merge.  Some unrelated pairs satisfying the velocity cut will evolve in such a way to satisfy the mass ratio limit.  By analyzing the full lightcone simultaneously, permitting relatively wide velocity offsets, we allow this to be satisfied across any cosmic time, increasing strongly the chances of getting a false positive pair.  

The second effect does reduce the tension identified in the previous subsection (from $\approx 8\sigma$ to $\approx 6\sigma$), but not enough to account for the entire discrepancy, because most of our pairs are selected prior to the merger stage when halo-finding effects merit the $t_{max}$ criterion.  As we show in Figure~\ref{fig:rates_tmax}, strong evolution in the observability time as $\tau = 4.8 (1+z)^{-2}$ provides again a satisfactory match between the pair and intrinsic merger rates with a $t_{\rm max}$ mass ratio definition.  

Therefore, we conclude that the mergers in the \illustris\ project have time-varying average observability time, in contrast to expectations based on simulations of isolated mergers \citep[e.g.,][]{lotz08,Lotz2010,lotz10} and the most common assumption in the survey literature \citep[e.g.,][]{Williams2011,Man2012,Man2016}.  In Section~\ref{s:properties} we analyze the characteristics of the simulated merging pairs that might give rise to such evolution.

\subsection{Comparison to Data} \label{ss:datacompare}

{ 
In Figure~\ref{fig:obscompare}, we compare our lightcone-based Illustris merger fraction estimates to observational measurements by \citet{Man2016}.  Here we use the same stellar mass ($M_* > 10^{10.8} M_{\odot}$) and separation ($10 < d / (\rm kpc/$$ h)< 30$) criteria for the simulated points as \citet{Man2016} used for the observed ones.  We combine the observational data for both major and minor pairs, and show the same stellar mass ratio selection ($M_{\rm sec}/M_{\rm pri} > 0.1$) for Illustris. 

For the simulated pairs, we applied a somewhat idealized selection in redshift space. Observationally, \citet{Man2016} used photometric redshifts, which have considerable uncertainties.  While we used an identical redshift difference selection for our pairs ($\Delta z < 0.2 (1+z)$) at most redshifts, for simplicity we did not include redshift uncertainties in our lightcone catalogs. Indeed, the incidence of false positive pairs in the simulation is about half that inferred statistically by \citet{Man2016} for similarly selected pairs with real photometric redshift errors. This difference causes the Illustris raw pair fractions to be somewhat lower (factor of $\sim 1.5$) than the uncorrected pair fractions in these data. Therefore, instead of comparing pair fractions, in Figure~\ref{fig:obscompare}, we compare the Illustris true merger fractions to the statistically corrected merger fractions quoted by \citet[][Table 1]{Man2016}.

In Figure~\ref{fig:obscompare}, we find reasonably good agreement between simulated and observed minor+major merger fractions.  At $z \lesssim 1$, the Illustris merger fractions are below the observed ones by about $\sim 50\%$, but this is near the uncertainty level, especially considering the large cosmic variance we find for the relatively small simulated lightcones at $z < 1$.  We find that the simulated and observed merger fractions are very similar at $z > 1$, with merger fractions $f_m \sim 0.15$ over $1 < z < 3$ \citep[see also:][]{Qu2016}.  Moreover, both observed and simulated mass-selected merger fractions are nearly constant over this time period of more than three Gyr. The simulated major+minor merger fraction also shows a hint of being roughly the same or a little higher ($f_m \sim 0.2$) at $3 < z < 5$, somewhat contrary to hints from current observational estimates that the merger fractions turn over. However, the theory and data points both are highly uncertain and/or incomplete at $z > 2.5$.  

This rough agreement between mock catalogs and observations supports our conclusion: in order to infer the true intrinsic merger rates (Figure~\ref{fig:rates}) from measurements of the merger or pair fraction, we probably must assume that the merger-pair observability time evolves to smaller values at high redshift.

\begin{figure}
\begin{center}
\includegraphics[width=3.5in]{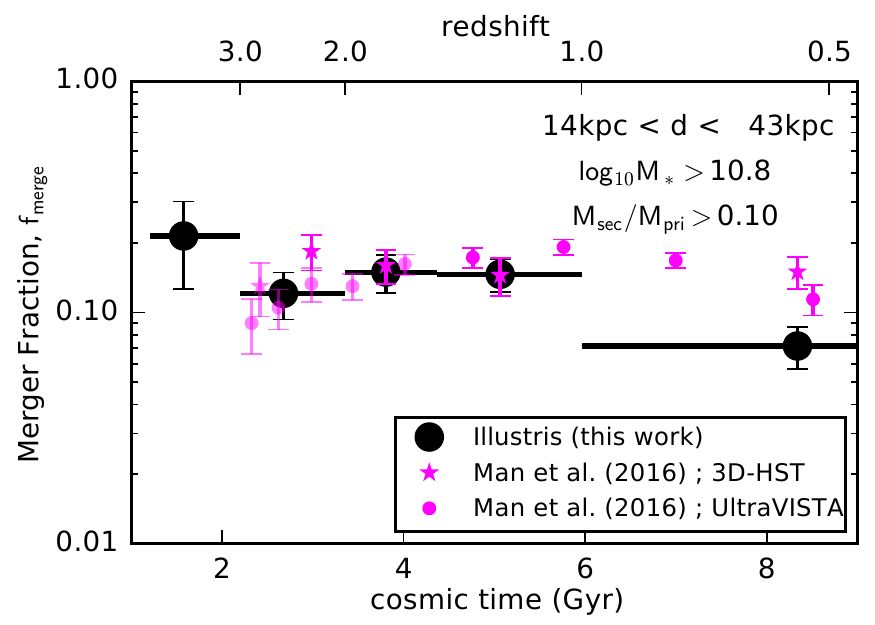}
\caption{   Here we compare Illustris mass-ratio-selected major+minor merger fractions to observational estimates, finding reasonably good agreement. We plot the 3D-HST+CANDELS-based and UltraVISTA-based merger fractions measured by \citet[][Table 1]{Man2016}, and show Illustris merger fractions using the same stellar mass ($M_* > 10^{10.8} M_{\odot}$), separation ($10 < d / (\rm kpc/$$ h)< 30$), and stellar mass ratio ($M_{\rm sec}/M_{\rm pri} > 0.1$) definitions.  While there is minor disagreement at $z < 1$, this could result from the much smaller simulated survey area compared with the observations. Moreover, the simulation displays similar behavior as observations at $z > 1$: the merger fraction is roughly constant over $1 < z < 3$. This behavior occurs despite a steeply rising intrinsic merger rate versus redshift (R-G15). \label{fig:obscompare} }
\end{center}
\end{figure}

}

\section{Evolution in Pair Properties}   \label{s:properties}

It may be natural to expect evolution in the merger-pair observability timescale.  For example, the linear-regime dynamical timescale is inversely proportional to the Hubble timescale $H(t)^{-1}$ at a given epoch, approximating $\tau \propto (1+z)^{-3/2}$, which is statistically consistent with our findings in Section~\ref{s:pairs}.  However, if this is driving a difference in the merger-pair observability times, then this signature should be imprinted on the properties of these merger events.  For example, the mergers might have a different distribution of velocities, orbits, or durations at early times.  

\begin{figure}
\begin{center}
\includegraphics[width=3.5in]{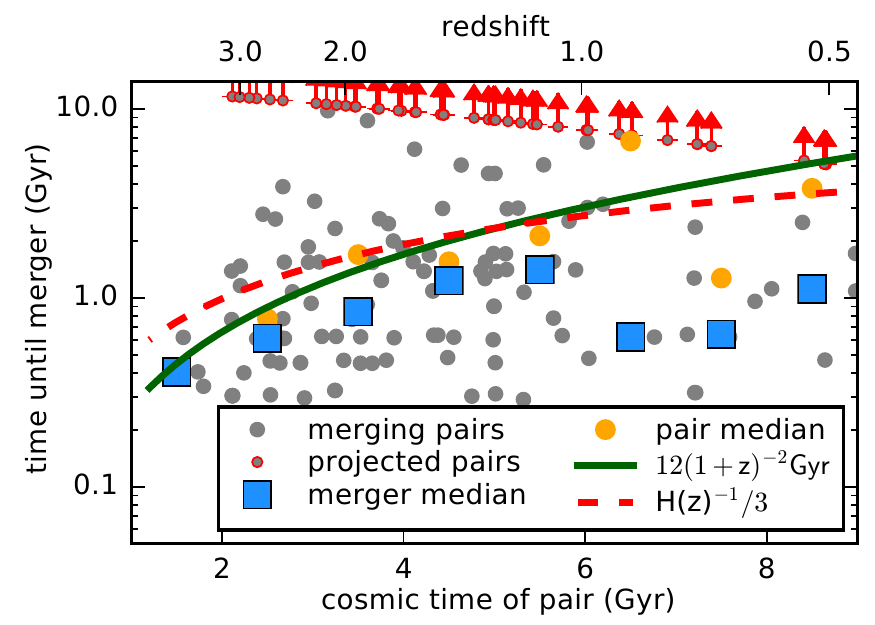}
\caption{ For each pair, we estimate the time until the pair merges, defined as both members having the same descendant at a given timestep from the merger trees.  For major merging pairs at $z > 1$, there is a clear median trend consistent with $(1+ z)^{\alpha}$ with $\alpha \sim -1$-$2$. At $z < 1$, we lack statistics to infer any possible trend. { Gray circles with red borders and arrows correspond to false positives --- massive pairs meeting our selection but failing to merge by $z=0$ --- so we plot their values as a lower limit, representing the time elapsed between the epoch at which we observed them and the present.} \label{fig:tmerge}}
\end{center}
\end{figure}

Figure~\ref{fig:tmerge} presents suggestive evidence that the merger event timescales evolve in a manner consistent with an evolving observability time.  At $z > 1$, there is clear evolution in the median time between when a pair is observed and when the final merger remnant forms, albeit with significant scatter ($\sim 0.3$ dex).  This trend is consistent with $(1+z)^{-2}$ or $(1+z)^{-3/2}$ at $z > 1$, but it is difficult to conclude anything at $z < 1$ owing to poor statistics.

\subsection{Consequences for remnant formation times}  \label{ss:remtimes}

\begin{figure}
\begin{center}
\includegraphics[width=3.5in]{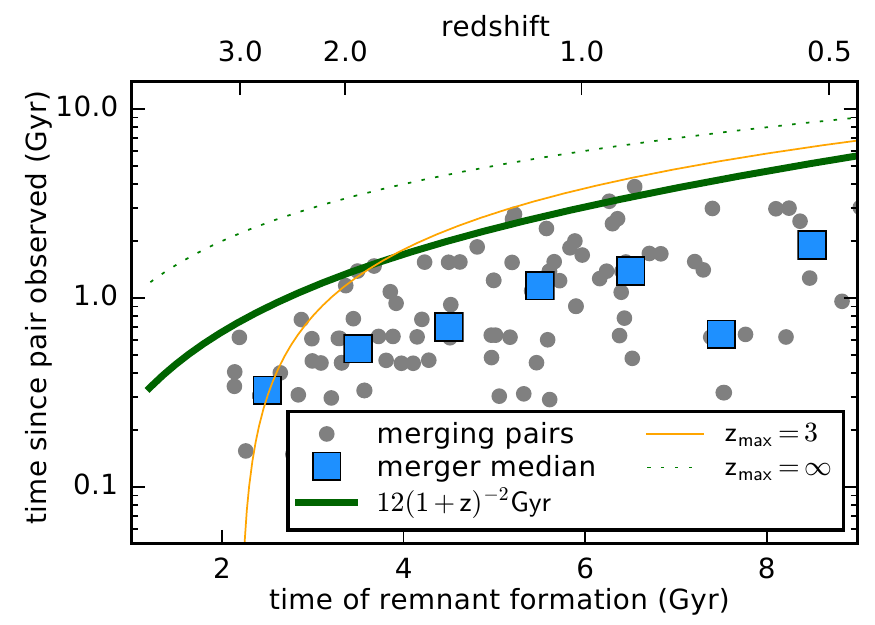}
\caption{ For each merging pair, we compute the time of merger (x axis), and plot the time since we selected the pair in our lightcone catalog.  On average, mergers at $z\sim 1$ were observed as pairs $\sim 1$ Gyr prior, while mergers at $z\sim 2$ were observed as pairs $\sim 0.5$ Gyr prior. Thus high-z pairs ``pile up'' as mergers at early times, boosting the merger rate early relative to late. We also plot the time since $z_{\rm max}=3$ and $z_{\rm max}=\infty $, which indicate the theoretical maximum curves for hypothetical surveys with some distance or redshift limitation.  \label{fig:tremnant}}
\end{center}
\end{figure}

A critical conceptual difference between a pair-defined merger rate and common intrinsic measurements is that the latter typically measure the \emph{instantaneous remnant formation rate} defined at the time the merger concludes, while pairs by necessity only measure merging systems at some time before the merger concludes and a remnant is formed.  Figure~\ref{fig:tremnant} shows a consequence of the evolving pair-merger timescale (Figure~\ref{fig:tmerge}), which also results from any pair survey with a fixed upper distance limit.  As a function of merger completion time, there is strong evolution in the median time since the pair was observed in the mock survey.  

In other words, for pairs that merge at $z = 1$, we observed them as pairs $\sim 1$ Gyr prior, on average.  However, for pairs that merge at $z \ge 2$, we observed them as pairs only $\sim 0.5$ Gyr prior, on average.  

Even if the median merger times do not evolve as in Figure~\ref{fig:tmerge}, the existence of any fixed upper distance limit in the survey will cause a qualitatively similar trend to occur, as shown by the dotted line and solid orange line in Figure~\ref{fig:tremnant}.  Given an unevolving but wide distribution of $t_{\rm merge}$, the upper tail of this distribution will merge much later than the lower tail.  As a function of the final remnant formation time, there will be at least minor evolution in the median time since the pairs were observed, with a trend similar to $H(t)^{-1}$, which we plotted in Figure~\ref{fig:tmerge}.

Also, this effect spreads out the merger events observed as pairs at high redshift into a range of future remnant formation times.  In Figure~\ref{fig:rates}, we track the volume density of these remnant formation events and the volume density of galaxies, plotting the ratio of these quantities divided by each time bin's duration as red triangles: \[ \frac{dN_{\rm mergers}}{dt} = \frac{n_{\rm mergers}} {n_{\rm gal} \Delta t}. \]   This calculation is very similar to the one used to measure the instantaneous merger rates themselves in the simulation (R-G15) and therefore we recover the expected strong redshift evolution in the red triangles of Figure~\ref{fig:rates}.  

\subsection{Rapid stellar mass growth}   \label{ss:tsfr}

While Figures~\ref{fig:tmerge} \& \ref{fig:tremnant} show tentative evidence for evolution in the merger event timescales for close pairs, this quantity is not used in the translation between pair fractions and merger rates in Figures~\ref{fig:rates} \& \ref{fig:rates_tmax}.  Instead, the relevant quantity is the merger-pair observability timescale, which we denote $\tau_{\rm obs}$.  This timescale encodes all factors that affect the translation between events of interest (final mergers of a given mass and mass ratio) and observed quantities (pairs of a given mass and mass ratio), including geometric effects owing to the projected distance cut, the average merger orbital properties and velocities, and stellar mass evolution which cause them to be selected or not by standard cuts.  

We investigated the average orbital properties of the mergers identified in Figure~\ref{fig:tmerge} by evaluating the Keplerian pericentric distance $r_{\rm peri}$ associated with the pair's current masses, positions, and velocities.  We find no systematic time evolution in average $r_{\rm peri}$, although we recover the expected behavior that pairs with smaller $r_{\rm peri}$ have much shorter times before the merger occurs.  This suggests that there is no major evolution in the average orbits of the simulated merging pairs at $1 \lesssim z \lesssim 3$.  Therefore, we do not expect that our projected distance or line-of-sight velocity cuts are selecting mergers at different stages at different epochs and causing the evolving observability time $\tau$.  

However, another factor may be driving a tendency to select mergers at a later relative stage in the early universe:  stellar mass growth.  We selected pairs having $ 10^{10.5} \le \log_{10} M_{\rm pri}/M_{\odot} < 10^{11}$ and $M_{\rm sec}/M_{\rm pri} > 0.25$.   On average, in both real galaxies and \illustris, the star formation rates (SFR) and specific star formation rates (sSFR) SFR$/M$ evolve very strongly with redshift for massive galaxies \citep{Noeske:2007a,Whitaker2012,Torrey2014}.  Therefore, a typical galaxy selected at $z = 1$ will have a much lower sSFR than another selected at $z = 3$, implying that the lower-redshift galaxy satisfies the mass cut for a longer period commensurate with its lower sSFR.  

\begin{figure}
\begin{center}
\includegraphics[width=3.5in]{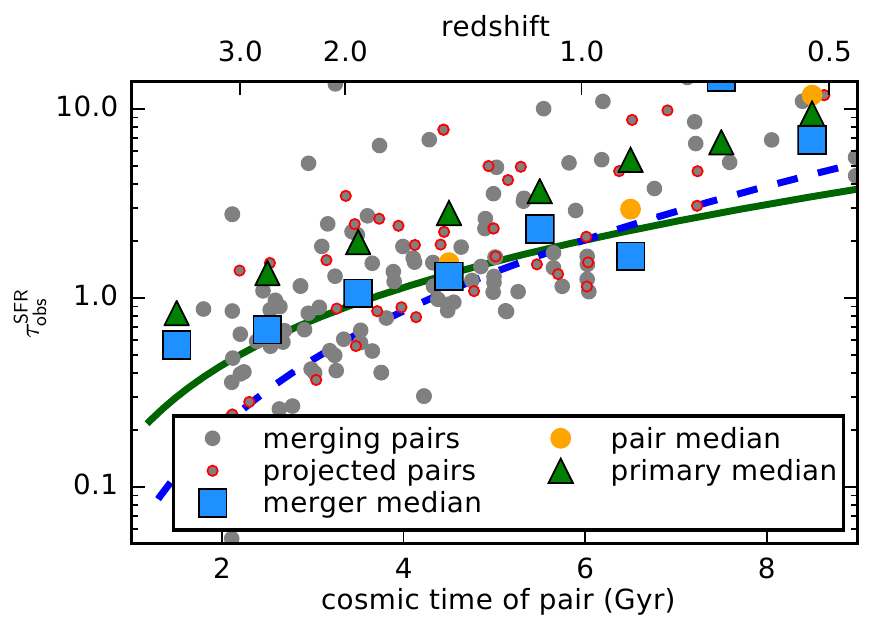}
\caption{ Here we estimate the time spent by merging pairs in our mass selection region.  We forward- and backward-model the current instantaneous SFR of the pair members and determine the total time that we  expect the pairs to have $10.5 < \log_{10} M_{\rm pri} < 11.0$ and $M_{\rm sec}/M_{\rm pri} > 0.25$.  The rapid decline in average sSFR, and total stellar mass growth, among all galaxies versus time causes galaxies at late times to spend much more time in the mass selection region.  \label{fig:tsfr} }
\end{center}
\end{figure}

We demonstrate this effect in Figure~\ref{fig:tsfr} by forward- and backward-modeling each galaxy's current SFR to estimate the duration each pair satisfies our mass selection criteria, $\tau^{\rm SFR}_{\rm obs}$.  As an example, suppose a selected pair evolves in such a way that its mass ratio is constant, and that the primary has SFR of $\dot{M}_*$.  We choose stellar mass limits $M_{\rm lower}$ and $M_{\rm upper}$. Then, the maximum mass-observability time is estimated as \[ \tau^{\rm SFR}_{\rm obs} = (M_{\rm upper} - M_{\rm lower})/\dot{M}_* . \]  Here we assumed a constant individual SFR, which neglects evolution in individual mergers but captures the average effect, and this approach has the advantage that $\tau^{\rm SFR}_{\rm obs}$ can be measured directly in observations.  This represents a lower limit to the stellar mass growth, which also has contributions from mergers themselves \citep[e.g.,][]{Rodriguez-Gomez2016}.  If this accretion is also significant, then the effect demonstrated in this section would be even stronger.

Figure~\ref{fig:tsfr} shows strong evolution in the estimated average time that merging pairs satisfy our mass selection criteria, from $\sim 2$ Gyr at $z=1$ to $\lesssim 1$ Gyr at $z=3$, following roughly $(1 + z)^{-2}$ in this redshift range.  This ignores the fact that mergers may or may not satisfy the velocity and distance cuts during their entire evolution, and so this timescale is an overestimate of the total observability times and the total merger event timescales.  However, if the average orbits do not evolve strongly (as we find), then we would expect these geometric effects to reduce the merger observability times by a constant factor, on average, and therefore $\tau_{\rm obs}$ may be proportional to $\tau^{\rm SFR}_{\rm obs}$.  

\subsection{Local environments of massive pairs}   \label{ss:delta}

\begin{figure}
\begin{center}
\includegraphics[width=3.5in]{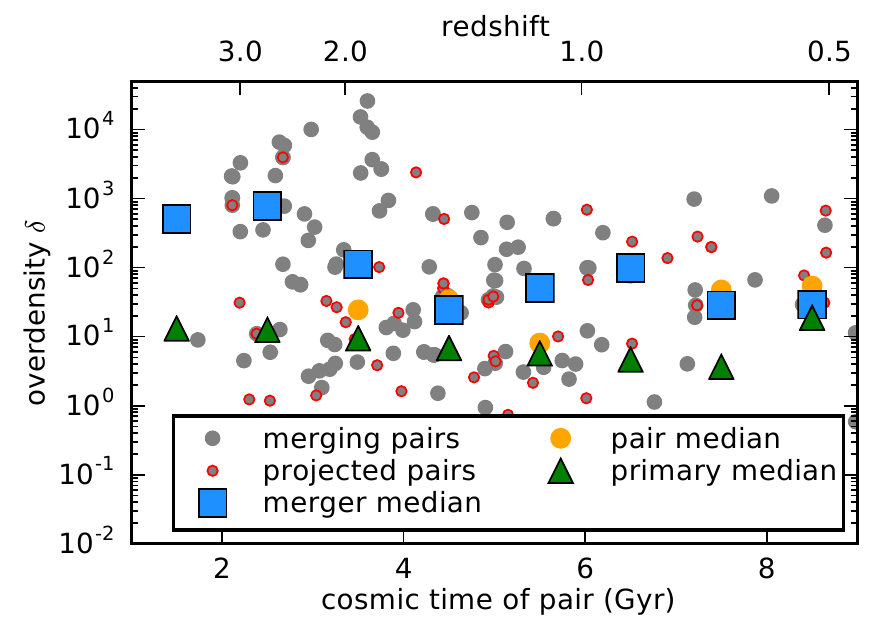}
\caption{ Overdensity of pairs versus cosmic time.  Pairs and mergers at $z \gtrsim 1$ reside in much denser regions relative to the typical primary massive galaxies at the same epoch. We do not find evidence that the time until merger in pairs (e.g., Figure~\ref{fig:tmerge}) depends on local density.  \label{fig:delta} }
\end{center}
\end{figure}

In Figure~\ref{fig:delta}, we plot the local overdensity $\delta$ of each pair versus cosmic time.  We define $\delta$ with a technique based on the three-dimensional distance to the fifth nearest tracer, similar to \citet{Vogelsberger2014b}.  First, the density field is smoothed by a three-dimensional Gaussian function with $\sigma$ equal to the distance to the 5th nearest tracer galaxy with r-band absolute magnitude $m_r \le -19.5$.  The quantity $1 + \delta$ is the ratio between the local number density of tracers and the average number density of tracers in the entire volume.  We plot the median $\delta$ value for merging pairs as blue squares, and for all (single and multiple) galaxies satisfying the same mass cut as green triangles.  

Merging (and all) pairs occupy regions that are roughly ten times denser than the average galaxy of the same stellar mass at $1 \lesssim z \lesssim 3$.  Our default selection criteria shows only tentative evidence for relative evolution in $\delta$ for merging pairs versus primaries from $z=1$ to $z=3$.  However, we find stronger evidence for such evolution when using selection criteria that yield more pairs (better statistics).  For example, using the same mass cut but a mass ratio cut of $0.1$ shows $\delta$ evolving from $\sim 2$ times greater in merging pairs ($\delta \approx 20$) to over $\sim 20$ times greater in merging pairs ($\delta \approx 300$) versus all galaxies.  Using the more complete $t_{\rm max}$ selection, with the default mass and mass ratio cuts, yields similar relative evolution.  Merging pairs with $M_{\rm pri} > 10^{10.8} M_{\odot}$ reside at $\delta \sim 10^3$ at $z\sim 2.5$  and $\delta \sim 100$ at $z \sim 1$, while the typical galaxy with $M > 10^{10.8} M_{\odot}$ has $\delta \sim 10$ across the whole redshift range.  

\begin{figure}
\begin{center}
\includegraphics[width=3.5in]{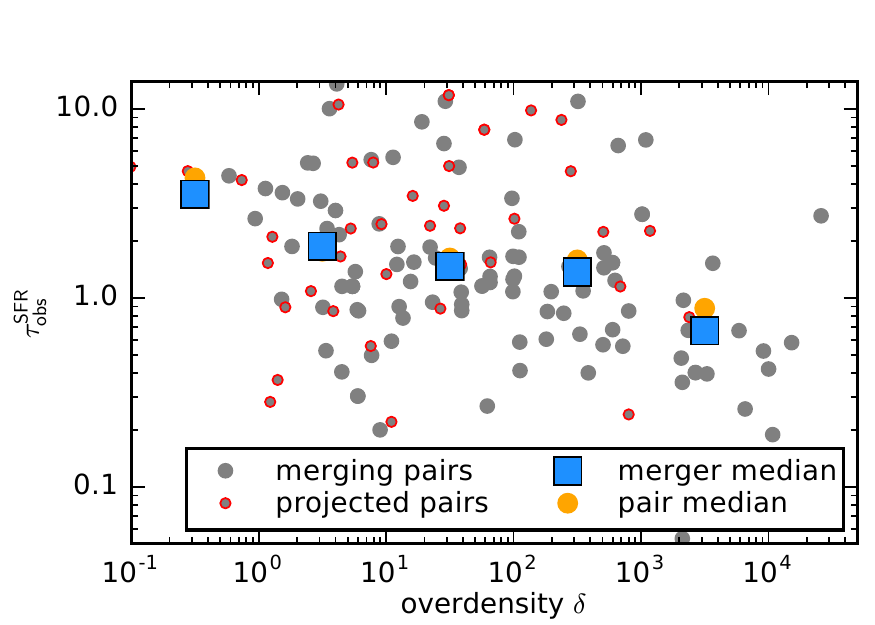}
\includegraphics[width=3.5in]{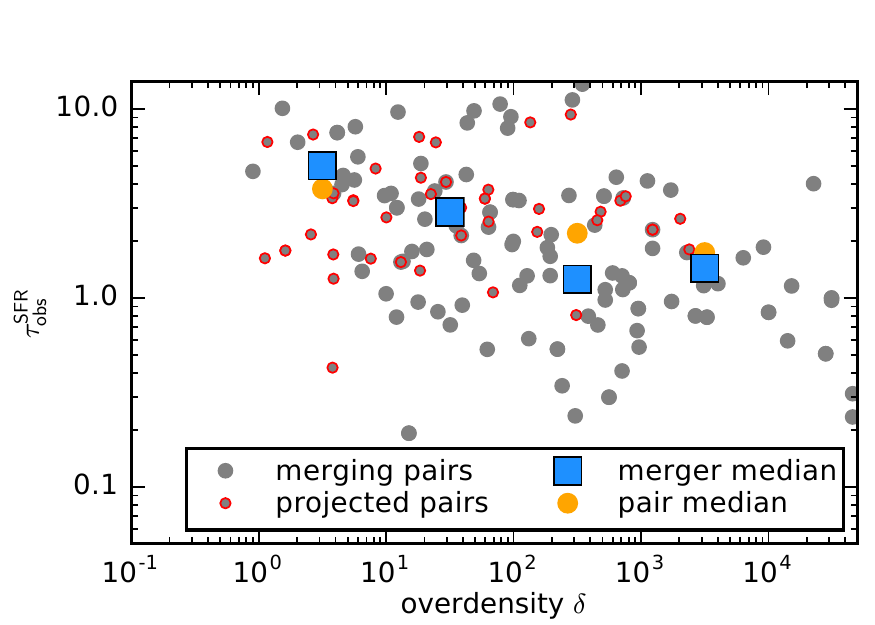}
\caption{ Star formation timescale $\tau_{\rm SFR}$ versus overdensity $\delta$.  Major pairs experience more in-situ stellar mass growth when they are in locally dense regions.  Top:  The default selection (Section~\ref{ss:catalogs}) used throughout this paper.  Bottom:  Same plot, except for higher stellar masses ($10.8 < \log M_{\rm pri}/M_{\odot} < 11.3$) and $M_{\rm sec} > 0.1 M_{\rm pri}$, confirming the tentative trend in the top panel. \label{fig:tsfr_delta} }
\end{center}
\end{figure}

Although we do not find that the times between observed pair and merger remnant depend strongly on $\delta$, Figure~\ref{fig:tsfr_delta} shows that $\tau^{\rm SFR}_{\rm obs}$ clearly does.  The effect is even more pronounced at higher stellar mass and lower mass ratios.  In other words, massive pairs in dense regions experience rapid in-situ stellar mass growth.  While not surprising, this effect could explain why massive major pairs at early times have shorter observability times and reside in dense regions.

\subsection{Timescales discussion}  \label{ss:discussion}

For mergers of a given pair of galaxies on the same orbits and the same total masses, the merger dynamical timescales $\tau_{\rm merge}$ will not differ with time.  But if the observable property (e.g., stars) of the halos changes differentially with time, then the observability time will change, all else equal.  

The central importance of the total observability time is shown when inferring the merger rate from pairs, as follows.  The number of intrinsic mergers per Gyr per galaxy is given by:
\[
R_m = \frac{ \phi_{\rm mergers}}{\phi} \frac{1}{ \tau_{\rm merge}},
\]
where $\phi$ are the galaxy volume densities and $\tau_{\rm merge}$ is the total merger timescale.  The observed pair fraction is given by:
\[
f_p = \frac{ \phi_{\rm mergers} }{\phi} \frac{\tau_{\rm obs}}{\tau_{\rm merge}},
\]
where $\tau_{\rm obs}$ is the observability time, which accounts for all selection effects. Then, to measure the merger rate from a pair fraction, we substitute:
\[
R_m = f_p / \tau_{\rm obs},
\]
independent of the intrinsic merger evolutionary timescale.

With perfect knowledge, a merger completing at time $t$ can be considered a merger for its entire past history, the age of the universe, $\tau_{\rm H (t)}$.  Any process that affects the time over which that merger would be selected will change $\tau_{\rm obs}$. This will only play a role if the observability-limiting timescale is shorter than the intrinsic merger event timescale.  Because galaxies evolve rapidly in SFR/M, $\tau_{\rm SFR} < \tau_{\rm H(t)}$ at $z \sim 2$, and therefore the stellar mass selection could limit $\tau_{\rm obs}$ at early times (Section~\ref{ss:tsfr}).  Then, at later times, dynamical friction could dominate, when $\tau_{\rm dyn} < \tau_{\rm H(t)}$, yielding a steady average $\tau_{\rm obs}$ as found by \citet{lotz08}.  In summary, to infer an event rate at a given time $t$, we should select the minimum of the possible observability-limiting timescales: $\tau_{\rm obs} = \rm min$$(\tau_{\rm H(t)}, \tau_{\rm dyn}, \tau_{\rm SFR})$.

\section{Discussion and Summary} \label{s:summary}

We analyzed the statistics of massive galaxy pairs in synthetic lightcones created from the \illustris\ simulation.  We found that mass-selected major pair fractions are roughly flat (Figure~\ref{fig:pairs}) as a function of cosmic time from $z\sim3$ to $z\sim1$, using a broad photometric redshift-inspired velocity cut of $\Delta z < 0.02 (1+z)$, for galaxies with $10^{10.5} M_{\odot} < M_{\rm pri} < 10^{11} M_{\odot}$ and $M_{\rm sec} > 0.25 M_{\rm pri}$. { In Figure~\ref{fig:obscompare}, we directly compared samples derived from our mock catalogs to measurements by \citet{Man2016}, finding good agreement in the value and evolution of the mass-ratio-selected merger fraction as a function of redshift.}  \citet{Qu2016} find similar trends from the Eagle Simulations, corroborating this result.  Because the intrinsic final merger rate of the same galaxies is a strong function of cosmic time (R-G15), and we showed that the false positive rate evolves only weakly (Figure~\ref{fig:mergers}), the merger-averaged observability time of pairs must evolve strongly with redshift (Figure~\ref{fig:rates}). On average, we find that mergers completing at early times were observed as pairs more recently than those at late times (Figure~\ref{fig:tremnant}).

Our results offer an explanation for the perplexing fact that mass-ratio-selected galaxy pair fractions level off at $z > 1$. If the observability times are lower than we assume, for any reason, then mergers are more common than we would infer by their number density divided by this constant timescale factor. In that case, our assessments of the importance of mergers in forming galaxies at early times would be incomplete. 

We found that the orbital parameters of the merging pairs we selected in the simulation do not evolve strongly over cosmic time, suggesting that changes in the merger pair observability time are not necessarily driven by changes of intrinsic  dynamical times. Therefore, other factors may be affecting the detectability of galaxy mergers as pairs in the early universe.

Observations have provided additional guidance about how to interpret mergers of very distant galaxy pairs.  For example, flux-selected major pair fractions evolve much more strongly with time \citep[e.g.,][]{Kartaltepe2007,Man2016,Mantha2016AAS}, more similar to theoretical expectations of mass-selected major pairs.  A possible interpretation is that total baryon mass might be a better discriminator of major pairs \citep[e.g.,][]{Man2016}, because secondary galaxies with high gas content might satisfy a flux ratio but not a mass ratio.  This could be an observational consequence of the scenario we outlined in Section~\ref{ss:tsfr}, whereby the rapid evolution in stellar mass growth by star formation implies an evolving observability time for mass-selected pairs.  In this scenario, rapid galaxy evolution in merging pairs (and all galaxies) is the factor driving the weak evolution in massive major pair fractions.

It should be possible to account for these effects in reconciling predicted and observed merger rates. Because we can often estimate the SFRs of each member of each pair, it may be possible to use a broad pair selection and estimate the ``final mass ratio'' of the pairs when they do merge. Combining this with predictions for the wide distribution of merger event types, star formation histories, and timescales, should allow us to make maximum use of our data to understand the varied evolutionary paths taken by mergers.

Our modeling approach utilized simulated galaxy catalogs, which were created using standard techniques for halo and galaxy finding in cosmological simulations. These techniques make various assumptions about how to assign mass to different galaxies, and therefore the resulting intrinsic merger rates can be strongly model-dependent (R-G15).  While we have selected merging pairs before these effects are the strongest, this situation is imperfect.  The ideal solution is to fully forward-model the simulated galaxy populations into the observational frame, and measure galaxy catalogs, masses, and mass ratios in the same way as the real sample to study.  Then we can compute the observed and simulated merger rate estimators in identical ways.

In principle, we can use large hydrodynamical simulations to produce mock observations with spatial resolutions comparable to space missions such as the \hubble, \webb, \euclid, and \wfirstfull, enabling us to also measure additional merger indicators, such as late-stage tidal features or multiple nuclei.  Initial efforts are underway to systematically mock-observe large sets of galaxies from hydrodynamical simulations \citep[e.g.,][]{Torrey2015,Snyder2015,Trayford2015,Trayford2016,Kaviraj2016}.  Because \illustris\ resolves galaxy structures on spatial scales below $\sim 1$ kpc at $z \gtrsim 1$, its mock survey fields contain sources appropriate for comparing with existing deep, high resolution, but narrow survey fields.  In preliminary work, we conducted a simple proof-of-concept to combine our Thin lightcone geometries with the spectral synthesis code \sunrise\ \citep{jonsson06,jonsson09}, and we will release these initial synthetic images to the community.  For example, \citet{Vogelsberger2014b} used these to compare the \illustris\ to the HST Ultra Deep Field.  Similarly, \citet{Kaviraj2016} presented a lightcone image from the Horizon-AGN simulation \citep[e.g.,][]{Dubois2014}. Therefore, in the future, large cosmological hydrodynamical simulations will support very useful public mock datasets following the examples of the Millennium Run Observatory (O13) and Theoretical Astrophysical Observatory \citep{Bernyk2016}.

\section*{Acknowledgements}
We thank the referee for suggestions that improved this paper.  We thank Shy Genel, Debora Sijacki, Mark Vogelsberger, Gerard Lemson, Chris Hayward, Benjamin Weiner, Rachel Somerville, Kameswara Mantha, Daniel McIntosh, Gurtina Besla, and Peter Behroozi for helpful discussions that contributed to this paper.  We thank Dylan Nelson for coordinating the \illustris\ data release.  GFS thanks the 2012 International Summer School for Astrophysical Computing (ISSAC2012) on astro-informatics hosted by the High Performance Astro-Computing Center (HIPACC) at the University of California San Diego Supercomputing Center.   GFS appreciates support from the \hst\ grants program, numbers HST-AR-$12856.01$-A and HST-AR-$13887.04$-A, as well as a Giacconi Fellowship at the Space Telescope Science Institute (STScI), which is operated by the Association of Universities for Research in Astronomy, Inc., under NASA contact NAS 5-26555.  PT acknowledges support provided by NASA through Hubble Fellowship grant HST-HF2-51384.001-A awarded by STScI.  This research has made use of NASA's Astrophysics Data System. Support for programs \#12856 (PI Lotz) and \#13887 (PI Snyder) was provided by NASA through a grant from STScI.  The figures in this paper were constructed with the Matplotlib Python module \citep{Hunter:2007}.  This work used Astropy \citep{Robitaille2013} extensively.

\bibliographystyle{apj}
\bibliography{$HOME/Dropbox/library}
\bsp


\end{document}